\documentclass[prl,twocolumn,floatfix,showpacs]{revtex4}
\usepackage{amssymb}
\usepackage{amsmath}
\usepackage{epsfig}
\usepackage{graphicx}

\begin{document}

\title {THE USE OF QUANTUM DOTS AS  HIGHLY
SENSITIVE SWITCHS BASED ON SINGLET-TRIPLET TRANSITION AND SYMMETRY
CONSTRAINT}

\author {Y.M. Liu}
\author {G.M. Huang}
\author {C.G. Bao*}

\affiliation{The State Key Laboratory of Optoelectronic Materials
and Technologies, and Department of Physics, Zhongshan University,
Guangzhou, 510275, P.R. China}

\begin{abstract}
Based on symmetry constraint that leads to the appearance of nodes
in the wave functions of 3-electron systems at regular triangle
configurations , it was found that, if the parameters of
confinement are skillfully given and if a magnetic field is tuned
around the first critical point of\ the singlet-triplet
transition, a 2-electron quantum dot can be used as \ a highly
sensitive switch for single-electron transport.
\end{abstract}

  \maketitle
* The corresponding author.

Modern experimental techniques, e.g., using electrostatic gates
and etching, allow a certain number of electrons ( from a few to a
few thousand) to be confined in quantum dots.$^{1-6}$\ \ As a kind
of systems different from those existing in nature, rich physics
is contained. \ Therefore, they attract certainly the interest of
academic research. \ On the other hand, the properties of the dots
can be controlled, e.g., by changing the gate voltage or by
applying an external magnetic field, etc. Therefore, these systems
\ have a great potential in application. \ In particular, the dots
can be used as a single-electron switch to control the electric
microcurrent. \ Making use of the Coulomb blockade $^{4,7}$,
current switches have already been designed by a number of authors
(refer to ref.8 and 9, and references therein). \ This kind of
efforts is crucial for developing microtechniques. \ In general,
there are two factors harmful to the sensitivity of this kind of
device, one is thermal fluctuation, another is quantum tunneling.
\ The former can be effectively reduced by lowering the
temperature, the latter can not. \ In order to eliminate the
quantum tunneling, we propose an idea of a new mechanism in this
paper. \ Related theoretical evaluation has been \ performed,
however, technical details are not involved.

\qquad Let us review briefly the main physics of Coulomb blockade. \ Let a
dot be connected to a source and a drain by tunnel barriers. \ A
gate-voltage $V_{g}$ is applied on the dot, while bias-voltages $V_{s}$ and $%
V_{d}$ \ are applied on the source and drain, respectively. Let \ $N$\
electrons be confined in the dot with the ground state energy $\epsilon _{N}$%
. \ If an extra electron comes in from the source, the lowest energy of the
(N+1)-electron system is denoted as $E_{N+1}$. \ The extra electron is
driven by the bias-voltage $V_{SD}=V_{s}-V_{d}$ . \ \ Thus the condition of
Coulomb blockade\ is \ $\Delta _{o}\equiv -e\beta V_{SD}<E_{N+1}-\epsilon
_{N}$ $\equiv \Delta $ (where the constant $\beta =0.81$ is introduced by
Foxman et al.$^{10}$ , \ however the actual value of $\beta $\ is not
essential to the following discussion). \ Whereas if $\Delta _{o}\geq \Delta
$, the blockade is released and the extra electron would go through the dot
to the drain. \ By increasing $V_{g},$ $\Delta $\ can be reduced as shown
schematically in Fig.1a. \ When $V_{g}$ increases and arrives at a critical
value, the condition $\Delta _{o}=\Delta $ holds, \ and resonant
transmission occurs. \ \ It is noted that $\Delta $ is changed continuously
with $V_{g}$. \ Therefore, even when $\Delta >\Delta _{o}$ , the leak of
current might occur due to quantum tunnelling. \ Thus, the blockade is not
strict, and the sensitivity of the device as a switch is more or less
spoiled.

\begin{figure} [htbp]
\centering
\includegraphics[totalheight=1.7in,trim=40 60 10 70]{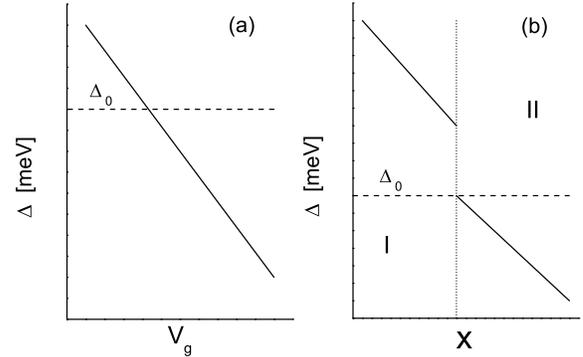}
\caption{A sketch to show the continuous variation of $\Delta $ in
the usual
device of Coulomb blockade\ (a), and in an improved device (b). \ When $%
\Delta =\Delta _{o}$, resonant transmission occurs. \ X in (b) is
a quantity to control the variation of $\Delta $ (not yet
specified)$\text{ } .\text{ }$} \label{Fig.1}
\end{figure}

\qquad\ \ On the other hand, if one can design a device so that $\Delta $
varies abruptly as shown in Fig.1b, the quantum tunnelling would be
remarkably suppressed and the sensitivity would be greatly improved. \ To
achieve the abrupt change, \ \ we suggest a source-dot-drain device with the
following three features.

\qquad (1) The dot  is axial symmetric and contains  two electrons. \ The
eigenstates states have the orbital angular momentum $L$ and total spin $S$
to be conserved. \ In particular, the ground state has $L_{2}$ , $S_{2}$ ,
and energy $\epsilon _{L_{2}S_{2}}$ \ .

\qquad (2) $V_{g}$ is fixed at a skillfully prescribed value (see below), \
while an adjustable external magnetic field $B_{o}$ is applied and tuned
around a critical point $B_{C}$ so that the ground state undergoes
singlet-triplet transitions$^{11-17}$. \ Accordingly, $(L_{2},S_{2})$ jumps
to $(L_{2}^{\prime },S_{2}^{\prime })$\ , or reversely, while the
corresponding ground state energies $\epsilon _{L_{2}S_{2}}$ \ and $\epsilon
_{L_{2}^{\prime }S_{2}^{\prime }}$ \ are identical at the critical point. \
For convenience, the region $B_{o}<(>)B_{C}$ is called region I (II).

\qquad (3) The bias-voltage is low and the extra electrons are controlled so
that they are coming slowly one-by-one, the incoming electron is so slow
that it is, relative to the center of the dot, in S-wave .

\ \qquad Let us see how this device works. \ When $B_{o}$\ is tuned in
region I, \ if the third electron can come in and form a 3-electron system,
then the quantum numbers of the 3-body final state are denoted as $%
(L_{3},S_{3})$ and the energy as $E_{L_{3}S_{3}}$. \ Once the partial wave
is limited to S-wave, 0nly $L_{3}=L_{2}$ and $S_{3}=S_{2}\pm 1/2$ \ are
allowed, accordingly $\Delta =E_{L_{3}S_{3}}-\epsilon _{L_{2}S_{2}}$ $\equiv
\Delta _{I}\;$(if more than one final states are allowed, the higher states
are not involved due to the low bias-voltage). \ When $B$ is in region II ,
since the quantum numbers of the 2-electron ground state has already been
changed,  $(L_{3},S_{3})$ \ are accordingly changed to $(L_{3}^{\prime
},S_{3}^{\prime })$ with the energy $E_{L_{3}^{\prime }S_{3}^{\prime }}$. \
Thus, after the transition, we have $\Delta =E_{L_{3}^{\prime }S_{3}^{\prime
}}-\epsilon _{L_{2}^{\prime }S_{2}^{\prime }}\equiv \Delta _{II}$.\ \
Although $\epsilon _{L_{2}S_{2}}=\epsilon _{L_{2}^{\prime }S_{2}^{\prime }}$
\ at the critical point, however in general \ $E_{L_{3}^{\prime
}S_{3}^{\prime }}\neq E_{L_{3}S_{3}}$, therefore $\Delta _{I}\neq \Delta
_{II}$ . \ This leads to a jump of\ $\Delta $ occurring at the critical
point as expected.

\qquad\ \ The realization of the abrupt variation of $\Delta $\
alone is not sufficient. \ In addition, two more conditions
$\Delta _{I}-\Delta _{II}>>0$ and $\Delta _{II}=\Delta _{o}$ are
required so that quantum tunnelling is nearly zero and $\Delta $
jumps down exactly to the right place to initiate the resonant
transmission as plotted in Fig.1b . \ It was found that, due to
the help of symmetry constraint as explained below, these two
conditions can be realized if the parameters of confinement are
specially designed. \ This is shown by a theoretical study of a
model as follows.

\begin{figure} [htbp]
\centering
\includegraphics[totalheight=1.9in,trim=20 40 10 40]{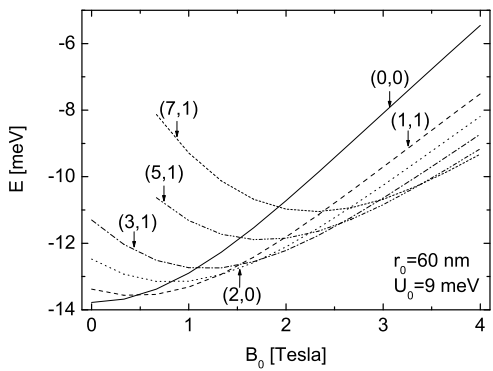}
\caption{Energies of the ground state of the 2-electron dot as a function of $%
B_{o}.$ \ The label $(L_{2},S_{2})$ \ is marked by the associated
curve$\text{ } .\text{ }$} \label{Fig.2}
\end{figure}

\vspace{1pt}\qquad\ For simplicity, \ the  model dot is assumed to be
2-dimensional. \ The potential of confinement \ is an axial-symmetric square
well \ $U(r)=-U_{o}$ if $\ r=\sqrt{x^{2}+y^{2}}\leq r_{o}$ , or \ \ $U(r)=0$
elsewhere, \ where \ $U_{o}$ is positive. \ Furthermore, a uniform \
magnetic field $B_{o}$ \ is applied perpendicular to the X-Y plane, \ The
Hamiltonian reads

$H=\sum\limits_{j=1}^{N}\;\lbrack -\;\frac{%
\hbar ^{2}}{2m^{\ast }}\nabla _{j}^{2}+U(r_{j})+\frac{1}{2}m^{\ast }(\frac{%
\omega _{c}}{2})^{2}\;r_{j}^{2}\rbrack -\frac{\hbar \omega _{c}}{2}\stackrel{%
\wedge }{L}-g^{\ast }\mu _{B}B\;S_{Z}\rbrack $ $+\frac{e^{2}}{4\pi
\;\varepsilon \;\varepsilon
_{0}}\sum\limits_{j<k}^{N}\frac{1}{r_{jk}}\ \ \ \ \ \ \ \ \ \ \ \
\ \ \ \ \ \ \ \ \ \ \ \ \ \ \  (1)$

where $\omega _{c}=eB/(m^{\ast }c)$ , \ $N$ is the number of electrons
(equal to 2 or 3), the Zeeman term is included. \ $m^{\ast }$=0.067$m_{e}$ ,
the dielectric constant $\varepsilon $=12.4 (for a GaAs dot), and the units $%
meV$ , $nm$ , and $Tesla$\ are used through out this paper. \ The
Hamiltonian is diagonalized to obtain the eigenenergies by using the methods
as outlined in \lbrack 18, 19\rbrack .

\begin{figure} [htbp]
\centering
\includegraphics[totalheight=1.6in,trim=40 30 10 40]{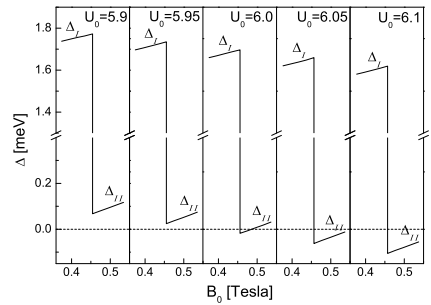}
\caption{$B_{C}$\ , the critical point of singlet-triplet
transition, as a function of \ $U_{o}$\ when $r_{o}$\ \ is fixed
(for a GaAs dot)$\text{ } .\text{ }$} \label{Fig.3}
\end{figure}

\qquad\ \ The ground state energies $\epsilon _{L_{2}S_{2}}$ calculated with
\ $U_{o}=9$ and $r_{o}=60$ (just as an example) are plotted in Fig.2 as a
function of $B_{o}$. \ The qualitative feature of Fig.2 would remain
unchanged if \ $U_{o}$ and $r_{o}$ vary inside a large reasonable region. \
One can see that the increase of $B_{o}$ causes a number of singlet-triplet
transitions and totally three critical points appear. \ It was found that
the vicinity of the first critical point, denoted as $B_{C}$ ( equal to $%
0.463$ in Fig.2) is suitable for our purpose. \ We shall concentrate in the
vicinity of $B_{C}$. \ In region I \ ($B_{o}<B_{C}$ ) and II, ($L_{2},S_{2})$
= $(0,0)$ and (1,1), respectively. \ It was found that, when $r_{o}$ is
given, the value of $B_{C}$ depends on $U_{o}$ nearly linearly. \ E.g., when
$r_{o}=35,\;B_{C}\approx 1.185+0.0285U_{o};$\ \ when $r_{o}=60,\;B_{C}%
\approx 0.325+0.0153U_{o}$\ . \ The larger the $U_{o}$, the larger the $%
B_{C} $, while the larger the $r_{o}$, the smaller the $B_{C}$.

\qquad When $B_{o}$ is given in region I, if the third electron with S-wave
can come in, ($L_{3},S_{3}$) should be (0,1/2). \ Selected values of $\Delta
_{I}=E_{0,1/2}-\epsilon _{0,0}$ are calculated and plotted as functions of $%
B_{o}$ as shown in Fig.3.\ \ On the other hand, from dynamical
consideration, if the three electrons form a regular triangle (RT), the
total potential energy can be minimized. Thus the RT is a favorable
configuration. \ However, from the study of symmetry constraint $^{20-22}$,
\ it was known that \ the spatial wave function of a 3-electron state $\Psi
_{LS}(123)$ would be zero at the RTs if $L=3j$ and $S=1/2$, \ or if $L\neq 3j
$\ and $S=3/2$, \ where $j$\ =0, 1, 2, $\cdot \cdot \cdot $. \ This is
called a prohibition of regular triangle (PRT), and will definitely cause
serious effect.

\qquad\ \ Let us prove the PRT when $S=3/2$. \ In this case $\Psi _{LS}(123)$
is totally antisymmetric. \ When the three electrons form an RT, a rotation
of the system by $\frac{2\pi }{3}$ is equivalent to a cyclic permutation of
particles. \ The rotation leads to an extra factor $e^{-i2\pi L/3}$\ , while
the permutation is equivalent to two interchanges and thus causes nothing. \
Thus, the equivalence leads to

$(e^{-i2\pi L/3}-1)\Psi _{LS}(123)=0\qquad \qquad \qquad \qquad
\qquad (2)$

This equation holds only at the RT configurations. \ Evidently, when $L\neq
3j$, the first factor of (2) is nonzero, therefore $\Psi _{LS}$ must be zero
at the RT. \ It implies the appearance of an inherent node at the RT, and
the PRT\ holds. \ The case $S=1/2$ can be similarly proved$^{20}$.

\begin{figure} [htbp]
\centering
\includegraphics[totalheight=2.5in,trim=30 15 10 45]{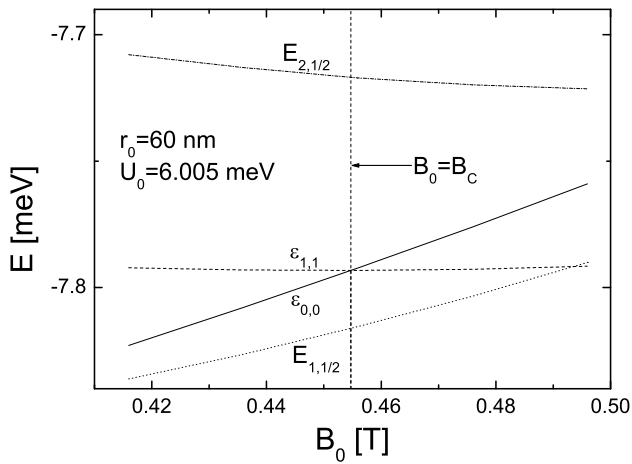}
\caption{The evolution of $\Delta _{I}$ and $\Delta _{II}$ \ with
respect to $B_{o}$ in the vicinity of \ $B_{C}$. \ \ $r_{o}$ is fixed at $%
60\;nm$ \ , and $U_{o}$ is given at a number of values (in
meV)$\text{ } .\text{ }$} \label{Fig.4}
\end{figure}

\qquad\ When PRT holds, the inherent node at the RT will remarkably increase
the kinetic energy, and at the same time push the wave function away from
the area of lower potential energy. \ Therefore, the associated energy $%
E_{L_{3},S_{3}}$\ is considerably higher than those free from the PRT. \ \
For this reason, $E_{0,1/2}$\ is in general high. \ By adjusting $U_{o}$ or $%
r_{o}$ , \ $E_{0,1/2}$ can be easily much higher than $\epsilon _{0,0}$\
resulting in a large $\Delta _{I}$ as confirmed by Fig.3.

\qquad In region II, the S-wave limitation leads to two choices ($%
L_{3},S_{3} $) = (1,1/2) and (1,3/2). \ Due to the PRT, \ $E_{1,3/2}$\ is
considerably higher and therefore can be neglected. \ \ Whereas, due to
being free from the PRT, \ $E_{1,1/2}$ is remarkably lower, \ this leads to
a small $\Delta _{II}=$ $E_{1,1/2}-\epsilon _{1,1}$as plotted in Fig.3. Thus
the PRT assures\ $\Delta _{I}-\Delta _{II}>>0$ , this leads to the great
jump shown in Fig.3.

\qquad\ The second condition $\Delta _{II}=\Delta _{o}$ can also be
satisfied  by adjusting the parameters of confinement. \ \ \ In what follows
$\Delta _{o}=0$ is assumed due to the very low bias-voltage. \ \ When $%
r_{o}=60$ , we know fom Fig.3  that $\Delta _{II}=\Delta _{o}=0$\ holds if $%
U_{o}=5.97$. \ In general, for a given $r_{o}$ , there is a corresponding $%
U_{o}=(U_{o})_{a}$ so that the pair \ of parameters $(U_{o})_{a}$ and $r_{o}$
gives \ $\Delta _{II}=0$ at $B_{o}=B_{C}$.\ \ Based on our numerical results
($30\leq r_{o}\leq 70$), a nearly linear relation $(U_{o})_{a}\approx
20.253-0.238\;r_{o}$ was found. \ The larger the $r_{o}$, the smaller the \ $%
(U_{o})_{a}$. \ When other types of confinement (e.g., a parabolic potential
with a limited height) are used, special set of parameters can also be found
to assure \ $\Delta _{II}=\Delta _{o}=0$ at $B_{o}=B_{C}$\ (not shown here).
Thus we can conclude that, if the parameters of confinement are skillfully
chosen, due to the symmetry constraint and making use of the singlet-triplet
transition, when $B_{o}$ is tuned in the vicinity of its first critical
point ($B_{C}$ ), the two-electron dot can work as a highly sensitive switch
at very low temperature. \ The regions I and II are associated with ''off''
and ''on'', respectively.

\qquad For a further discussion, a typical low-lying spectrum of the 2- and
3-electron dots in the vicinity of $B_{C}$ is given in Fig.4 (where $U_{o}$
is a little larger than $(U_{o})_{a}$). \ For these systems the levels not
appearing in the figure are much higher. \ If $U_{o}$ is given considerably
larger than $(U_{o})_{a}$ , \ $E_{1,1/2}$ would become considerably lower
and lead to \ $\Delta _{II}<<0$. \ In this case the intruding electron \ \
may find some way to release its energy (e.g., by emitting a photon or a
phonon) and falls into the level $E_{1,1/2}$. \ If this occurs, the
intruding electron can not go out again but remain inside, thereby the
previous mechanism of transport would be spoiled. \ Therefore $U_{o}$ is
better given at $(U_{o})_{a},$ this leads to $\Delta _{II}=\Delta _{o}$. \
In this case a resonant transmission takes place and the intruding electron
has no chance to remain inside.

\begin{figure} [htbp]
\centering
\includegraphics[totalheight=1.8in,trim=20 50 10 50]{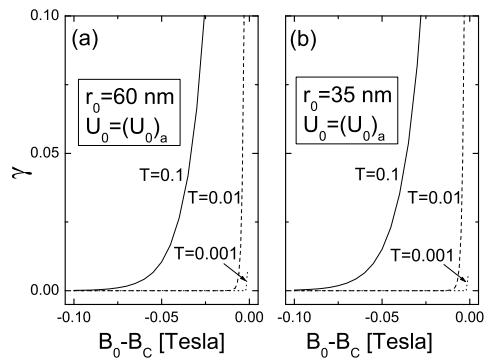}
\caption{The evolution of $(U_{o})_{a}$ versus \ $r_{o}$. $\ $The
parameters $(U_{o})_{a}$ and the associated \ $r_{o}$ together
assure the condition of resonant transmission, i.e., $\Delta
_{II}=\Delta _{o}$, (here $\Delta _{o}=0 $ is assumed) $\text{ }
.\text{ }$} \label{Fig.5}
\end{figure}

\vspace{1pt}\qquad If the confinement potential contains non-symmetric
component $U_{non}({\bf r})$ , $L$ is no more strictly conserved. \  In
region I (in the status ''off'')  if the 2-body ground state contains a
small  $L_{2}=1$ component, this component can absorb the incoming S-wave
electron and goes to the $E_{1,1/2}$ level,  thereby a leak may occur. \
Alternatively,  if the 2-body ground state contains a small $L_{2}=2$ \
component, this component can not  absorb the extra S-wave electron because
the level $E_{2,1/2}$ is quite high (cf. Fig.4). \ \ This fact implies that
only the kind of  $U_{non}({\bf r})$ with odd parity would cause a leak,
this kind  should be diminished as far as possible (e.g., in the technical
aspect, the two pipes connected to the dot are placed by the two sides as
symmetric as possible). \ \ \ A representative \ $U_{non}$ with odd parity
can be written as \ $U_{non}(r,\theta )=U(r)(\lambda _{a}\cos \theta
+\lambda _{b}\sin \theta )$, where $U(r)$ is the original one in eq.(1) .\
By using the perturbation theory ,  for two representative cases, the ratio
of the weights of the $L=1$ and $L=0$ components in the 2-body ground state
is $\tau =1.15(\lambda _{a}^{2}+\lambda _{b}^{2})$ if $r_{o}=60$, $%
U_{o}=(U_{o})_{a}=5.971$ and $B_{o}=B_{C}=0.416$; \ and $\tau =0.131(\lambda
_{a}^{2}+\lambda _{b}^{2})$ if $r_{o}=35$ , $U_{o}=(U_{o})_{a}=11.922$ , and$%
\ B_{o}=B_{C}=1.525$\ . \ From this evaluation, \ if $\lambda _{a}$ and $%
\lambda _{b}$ are in the order of 10$^{-1}$ , \ $U_{non}$ would not cause a
serious problem. \ Besides, a smaller $r_{o}$ would lead to a larger
level-spacing $\epsilon _{1,0}-\epsilon _{0,0}$, and therefore a smaller $%
\tau $ . \

\qquad\ \ \ The Kondo effect would do no harm in this device. \ This effect
ceases to exist in the status ''off''  because the spin of the dot is
meanwhile zero, while this effect is promotive in ''on''. \ \ However, the
thermal fluctuation might spoil the sensitivity of the device. \ \ In the
status ''off'', if the 2-electron state has been raised up from the ground
state to the level $\epsilon _{1,1}$ by thermal fluctuation, the third
electron may come in and the system may fall to the level $E_{1,1/2}$, and
thus the ''off'' is no more strict. \ Let us define$\ \gamma =\exp \{-($ $%
\epsilon _{1,1}-\epsilon _{0,0})/k_{B}T\},$ which is the ratio of the
probabilities of staying in the levels $\epsilon _{1,1}$ and $\epsilon _{0,0}
$ , respectively, and is plotted in Fig.5 as a function of $B_{o}$ , where $T
$\ is given at a number of values. \ From this figure, we know that, if $%
T\leq $ 0.01K, when $|B_{o}-B_{C}|>0.01,$ the thermal fluctuation is
negligible. \ Therefore, in this temperature, \ a change of $B_{o}$ in the
order of 0.01 around $B_{C}$\ is sufficient to initiate the ''on'' or
''off'' of the switch. \ When $T$ is much smaller than 0.01K, a very small
change of $B_{o}$ is sufficient to initiate the ''on'' or ''off'', thus the
switch becomes highly sensitive.

\qquad\ The S-wave limitation is a basic requirement. \ To evaluate in a
semi-classical way, let us assume that a classical electron comes in with a
velocity $\nu $ along a straight pipe with a radius $d_{c}$ aiming at the
center of the dot. \ Then the angular momentum is $m^{\ast }\nu d_{c},$ and
the S-wave limitation is $m^{\ast }\nu d_{c}\leq \hbar /2$\ , or $d_{c}\nu
<0.00288\;C$ , where $d_{c}$ is in $nm$, and $C$ is the velocity of light in
vacuum. \ \ On the other hand, the thermal velocity $\nu _{ther}=\sqrt{%
2k_{B}T/m^{\ast }}=0.0000710\sqrt{T}C$, where $T$ is the temperature in $K$.
\ \vspace{1pt}Thus, the requirement can be rewritten as $d_{c}<40.56(\nu
_{ther}/\nu )/\sqrt{T}$\ . \ If  $T=0.01,$ even $\nu $ is as large as $10\nu
_{ther}$\ , the requirement, namely $d_{c}<40.56,$\ can be realized with the
present technics of fabrication. \

\qquad As final remarks, two distinguished features of the suggested device
as a sensitive switch are noticeable. \ Firstly, the quantum tunneling has
been greatly suppressed. \ Secondly, the device is highly sensitive to the
variation of $B_{o}$ at low temperature, \ this is the superiority. \ \
However, since very low temperature together with the S-wave limitation and
good axial symmetry of the dot are required, the effectiveness of the switch
depends on the technical aspect, and remains to be checked. \ Nonetheless,
it is believed that the idea proposed in this paper, i.e., making use of the
effect of symmetry constraint and phase transition, is in general useful in
the design of various microdevice.

\section*{ Acknowledgment}
This paper is supported by the NSFC\ of China under the grant
No.90103028, 90306016, and No.10174098.
\vspace{1pt}

\end{document}